\documentclass[twocolumn]{aastex6}

\usepackage{multirow, graphicx, epsfig}
\DeclareGraphicsExtensions{.eps}

\slugcomment{Draft version \today}

\shorttitle{Characterizing the $\gamma$-ray Spectra of Blazars}
\shortauthors{van den Berg et al.}

\begin{document}

\title{Systematic Physical Characterization of the $\gamma$-ray Spectra of 2FHL Blazars}

\author{Jacobus P. van den Berg and Markus B\"{o}ttcher}
\affil{Centre for Space Research, North-West University, Potchefstroom, 2531, South Africa}
\email{24182869@nwu.ac.za, Markus.Bottcher@nwu.ac.za}
\and
\author{Alberto Dom\'{\i}nguez and Marcos L{\'o}pez-Moya}
\affil{Grupo de Altas Energ\'{\i}as and IPARCOS, Universidad Complutense de Madrid, E-28040 Madrid, Spain}
\email{alberto@gae.ucm.es, marcos@gae.ucm.es}

\begin{abstract}
We test different physically motivated models for the spectral shape of the $\gamma$-ray emission in a sample of 128 
blazars with known redshifts detected by the {\it Fermi} Large Area Telescope (LAT) at energies above 50~GeV.
The first nine years of LAT data in the energy range from 300~MeV to 2~TeV are analyzed in order to extend the spectral 
energy coverage of the 2FHL blazars in our sample. We compare these spectral data to four leptonic models for the 
production of $\gamma$-rays through Compton scattering by a population of electrons with different spectral shapes. In the 
first three models we consider Compton scattering in the Thomson regime with different acceleration mechanisms for the 
electrons. In the fourth model we consider Compton scattering by a pure power law distribution of electrons with spectral 
curvature due to scattering in the Klein-Nishina regime. The majority of blazar $\gamma$-ray spectra are preferentially 
fit with either a power law with exponential cut-off in the Thomson regime or a power law electron distribution with 
Compton scattering in the Klein-Nishina regime, while a log-parabola with a low-energy power-law and broken 
power-law spectral shape in the Thomson regime appears systematically disfavoured, which is likely a consequence of
the restriction to pure Thomson scattering which we imposed on those models. 
This finding may be an indication that the $\gamma$-ray emission from FSRQs in the 2FHL catalog is dominated by 
Compton scattering of radiation from the dusty torus, while in the case of BL Lac objects, it is dominated by synchrotron 
self-Compton radiation.
\end{abstract}


\section{Introduction}
\label{sec:Intro}

Blazars are a subclass of active galactic nuclei (AGN) whose jets are oriented at a small angle with respect to an 
observer's line of sight. This geometry leads to relativistic aberration effects and Doppler boosting along the jet 
direction. Blazars are characterized by strong non-thermal emission across the electromagnetic spectrum, rapid 
variability, and high optical polarization. This is the brightest and most numerous source class in the persistent 
extragalactic $\gamma$-ray sky \citep{aceroetal2015}.

The spectral energy distributions (SEDs) of blazars are characterized by two broad, non-thermal components. It is 
widely accepted that the low-energy component is due to synchrotron radiation (SR) of relativistic electrons (and 
possibly positrons) accelerated in the blazar jet. For the high-energy component, both leptonic and hadronic origins 
are possible \citep[e.g.,][]{bottcheretal2013}. In leptonic models, the X-ray and $\gamma$-ray emission is caused by 
inverse Compton (IC) scattering of low-energy photons by the same population of electrons which produced the SR. In 
this case, the shape of the $\gamma$-ray spectrum is directly related to the energy distribution of the accelerated 
electrons. This correlation is straightforward in the case of Compton scattering in the Thomson regime, but more 
complex in the Klein-Nishina regime \citep{bottcheretal2012, dermermenon2012}. Whether $\gamma$-ray production by
Compton scattering proceeds in the Thomson or Klein-Nishina regime, depends critically on the characteristic target 
photon energy. If the target photons originate from the co-spatially produced synchrotron emission (typically peaking
in the infrared to optical regime in the co-moving frame, leading to synchrotron self-Compton, SSC, emission) or from a 
dusty torus around the central accretion flow (with target photons in the infrared, leading to external Compton on dust 
torus emission), then the Compton scattering to GeV $\gamma$-ray energies typically occurs in the Thomson regime. In 
the case that the target photons originate externally from the Broad Line Region (dominated by optical to ultraviolet 
photons in the stationary frame of the AGN, leading to external Compton on BLR emission), then the Compton scattering 
to GeV energies typically occurs in the Klein-Nishina regime. A deviation of the $\gamma$-ray spectra of blazars from a 
pure power law may thus be caused either by an underlying electron population that deviates from a pure power law, 
and/or by the transition of the Compton scattering process from the Thomson to the Klein-Nishina regime towards higher 
$\gamma$-ray energies. 

Evidence for non-power law electron distributions has been found in the synchrotron continuum spectra of blazars. 
\citet{landauetal1986} showed that the low-energy peak of fifteen (out of a sample of eighteen) blazars are well 
fitted by a log-parabolic form. These authors showed that an energy dependent probability of stochastic acceleration, 
specifically if the acceleration probability decreases with increasing energy, leads naturally to an electron 
distribution with a log-parabolic form. In this context, the curvature of the spectra is not simply due to energy 
losses but is rather a direct consequence of the acceleration mechanism. This result was verified for the case of Mrk 
421 \citep{massaroetal2004, tramacereetal2007} and for other BL Lac objects \citep[][and references therein]
{tramacereetal2011}. In particular, \citet{massaroetal2004} also showed that a power law with exponential cut-off does 
not fit the synchrotron spectrum of Mrk 421 satisfactorily. This spectral shape might be expected if some limiting 
process is present in an acceleration mechanism such as diffusive shock acceleration
\citep[DSA; e.g.,][]{kirkheavens1989, ellisonetal1990, summerlinbaring2012}.

It has long been also recognized that the $\gamma$-ray spectra of blazars cannot be fitted by a simple power law 
\citep{abdoetal2010a}. This is expected in the framework of leptonic models, where the same electron population 
produces both the synchrotron and $\gamma$-ray emission through Compton scattering \citep[e.g.,][]{boettcher2007}. Note 
that the shape of the underlying particle distribution will determine the shape of the Compton $\gamma$-ray spectrum 
(see Section~\ref{sec:Model}).

Being able to characterise the high-energy spectra of a large sample of blazars may allow us to probe the underlying 
relativistic electron distribution and the characteristic energy of target photons for Compton scattering. Therefore, 
this methodology is a tool to diagnose the physical mechanisms of particle acceleration in the jets of blazars.

In this work, we compare the broad-band $\gamma$-ray spectra of 128 blazars selected from the Second Catalog of 
Hard {\it Fermi}-LAT Sources (the 2FHL catalog) with physically motivated models, over an energy range of almost four 
orders of magnitude, in an attempt to systematically characterize the spectral shape of the high-energy turnover. We 
stress that we do not aim to constrain physical parameters, but only investigate statistically the underlying physical 
processes.

The outline of this paper is as follows. Section~\ref{sec:Data} describes our blazar sample and data analysis. In 
Section~\ref{sec:Model}, the high-energy $\gamma$-ray spectra for the four theoretical models are derived. We describe 
our fitting methodology in Section~\ref{sec:Fitting} leading to the results presented in Section~\ref{sec:Results}. 
Finally, Section~\ref{sec:Summary} contains a summary and discussion of the results.

\section{Source Sample and Data Analysis}
\label{sec:Data}

In this Section, we describe our blazar sample and data analysis. Figure~\ref{fig:SampleSEDs} shows four examples of 
the spectral results from our analysis.

\subsection{Description of the Sample}
\label{subsec:SampleDescription}

Our sample includes all the 128 blazars with known redshifts from the {\it Fermi} Large Area Telescope (LAT) 2FHL 
catalog \citep[sources detected at energies larger than 50~GeV,][]{ackermannetal2016,dominguez2015}. The redshifts 
range from $z=0.004283$ (M87) to $z=2.1$ (MG4 J00800+4712), with the median of the distribution at $z=0.215$.

Blazars tend to be divided in two main populations according to properties of their optical spectra. There are (almost) 
featureless objects known as BL Lac blazars, and flat-spectrum radio quasars (FSRQs), typically characterized by broad 
emission lines \citep{urrypadovani1995}. According to the blazar sequence, which is empirically derived, BL Lacs are 
characterized, on average, by harder $\gamma$-ray spectra and lower luminosity than FSRQs \citep{fossatietal1998, 
ghisellinietal2017}. Our sample contains 106 BL Lacs (with or without prominent galaxy emission), 10 FSRQs, 4 blazars 
of uncertain type (BCUs), and some radio galaxies and other types of AGN.

Another blazar classification methodology is motivated by the frequency at which their synchrotron peak is located. 
This characteristic frequency, which is provided in the 2FHL, classifies these sources as low-synchrotron peak (LSP), 
intermediate-synchrotron peak (ISP), and high-synchrotron peak (HSP) blazars with their synchrotron peak frequency 
at $\log_{10}(\nu^{s}_{\rm peak})<14$, $14<\log_{10}(\nu^{s}_{\rm peak})<15$, $\log_{10}(\nu^{s}_{\rm peak})>15$, 
respectively, with $\nu^{s}_{\rm peak}$ given in units of Hz. The 2FHL blazars are mostly catalogued as HSP BL Lacs 
\citep[see Figure 8 in][]{ackermannetal2016}. The exact numbers in our sample are 33 LSP, 12 ISP, and 82 HSP blazars 
(there is one source without clear classification due to a poorly sampled SED).

\subsection{Data Analysis}
\label{subsec:DataAnalysis}

The first nine years (450 weeks, from MJD~56048 to MJD~57772) of {\it Fermi}-LAT data were analyzed in the energy 
range from 300~MeV to 2~TeV in order to extend the energy spectral coverage of the 2FHL blazars in our sample. We 
analyzed this data set using the P8R2\_SOURCE\_V6 instrument response functions and the {\it Fermi} Science Tools 
version v10r0p5. Events were selected within a circular region of interest (ROI) of $15^{\circ}$ centred at the 
2FHL source position. We selected ``Source'' class events (evclass = 128 and evtype = 3) that were recorded only when 
the telescope was in nominal science mode. 
To reject the background coming from the Earth's limb, we selected photons with a zenith angle $ \leq 90^{\circ}$. For 
the spectral reconstruction, a binned likelihood analysis was performed making use of the {\it pyLikelihood} python 
module of the {\it Fermi} tools. We started by including all the sources from the Third {\it Fermi} Source Catalog 
\citep[3FGL,][]{aceroetal2015} in the spectral-spatial model. All the 3FGL sources were assumed to have spectral types 
as suggested in the catalog. The spectral parameters for sources with a significance larger than $5\sigma$ and located 
less than $5^{\circ}$ away from the ROI centre were left free. We also let the normalization factor of the isotropic 
(iso\_P8R2\_SOURCE\_V6\_v06.txt) and Galactic (gll\_iem\_v06.fits) background models be free. For the rest of the 
sources all the parameters were left fixed to their catalog value. Finally, all sources with significance lower than 
$2\sigma$ were removed from the model. For the calculation of the spectral points, we repeated the procedure in each 
energy bin using a power law with the normalization factor free and the spectral index fixed to 2 (where the spectral 
index $\Gamma$ is defined as $\propto E^{-\Gamma}$). Whenever the significance of the spectral point was less than 
$1.5\sigma$, an upper limit was calculated instead.

\begin{figure*}[!ht]
\centering
    \includegraphics[width=8.2cm]{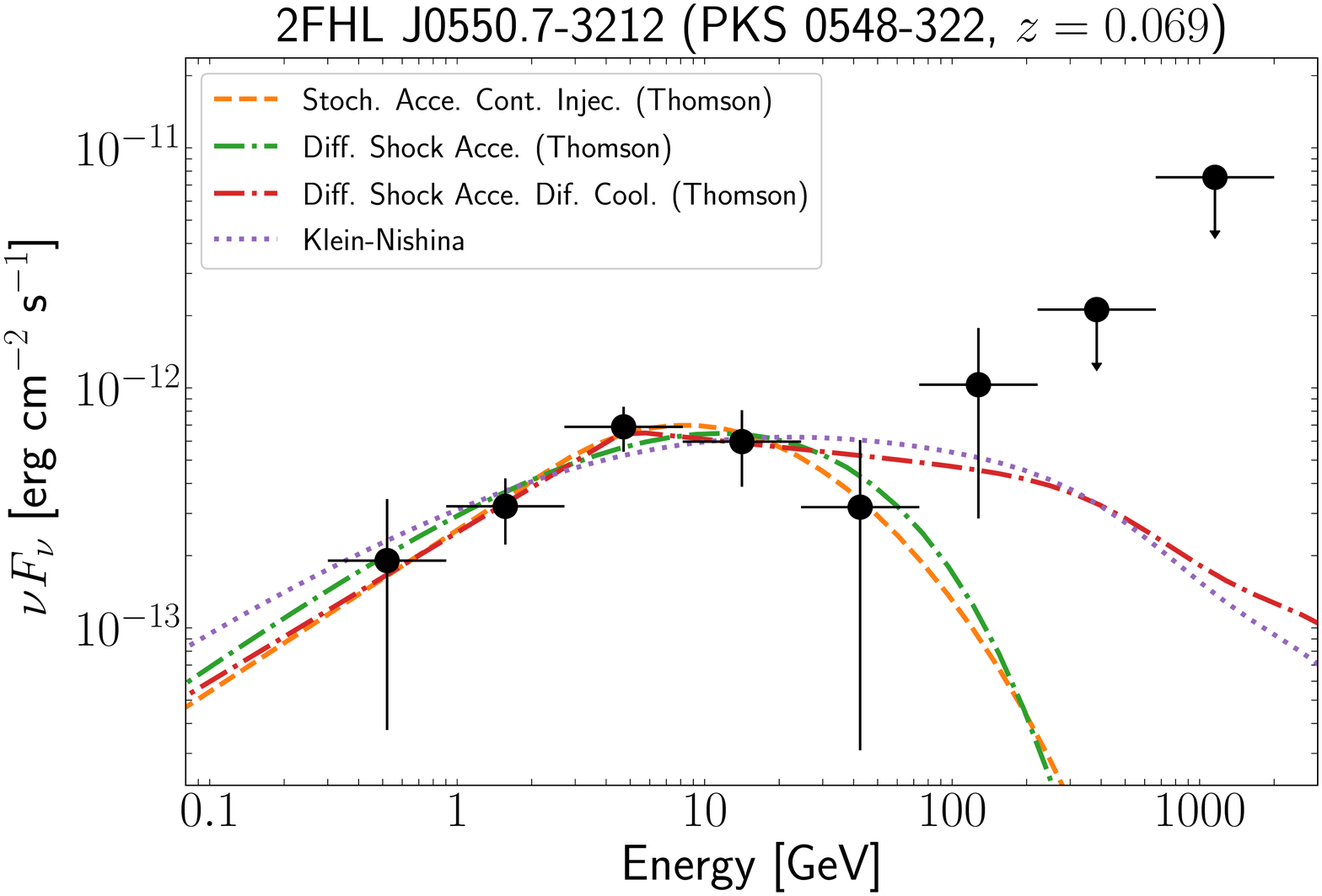}
    \includegraphics[width=8.2cm]{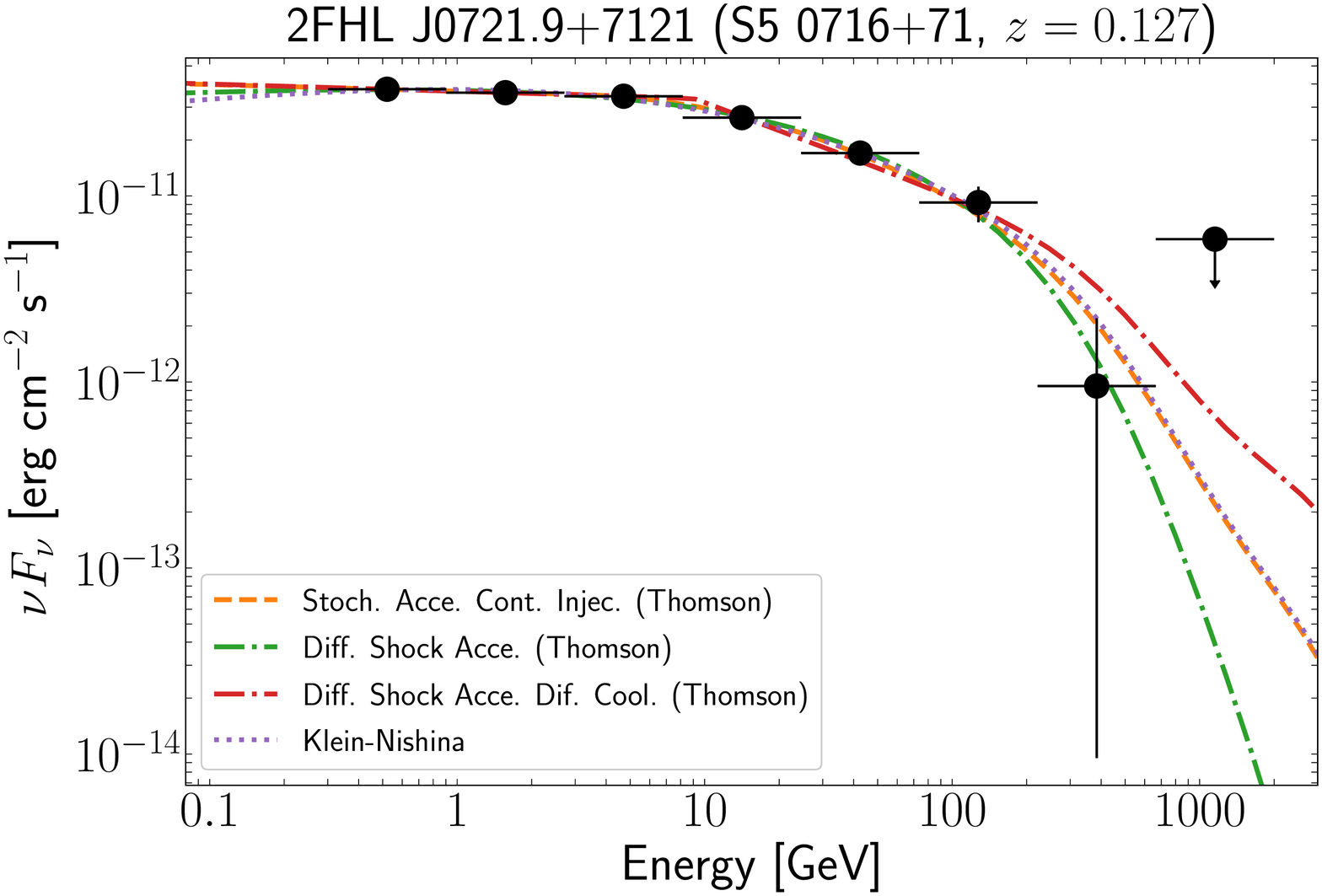}\\
	\includegraphics[width=8.2cm]{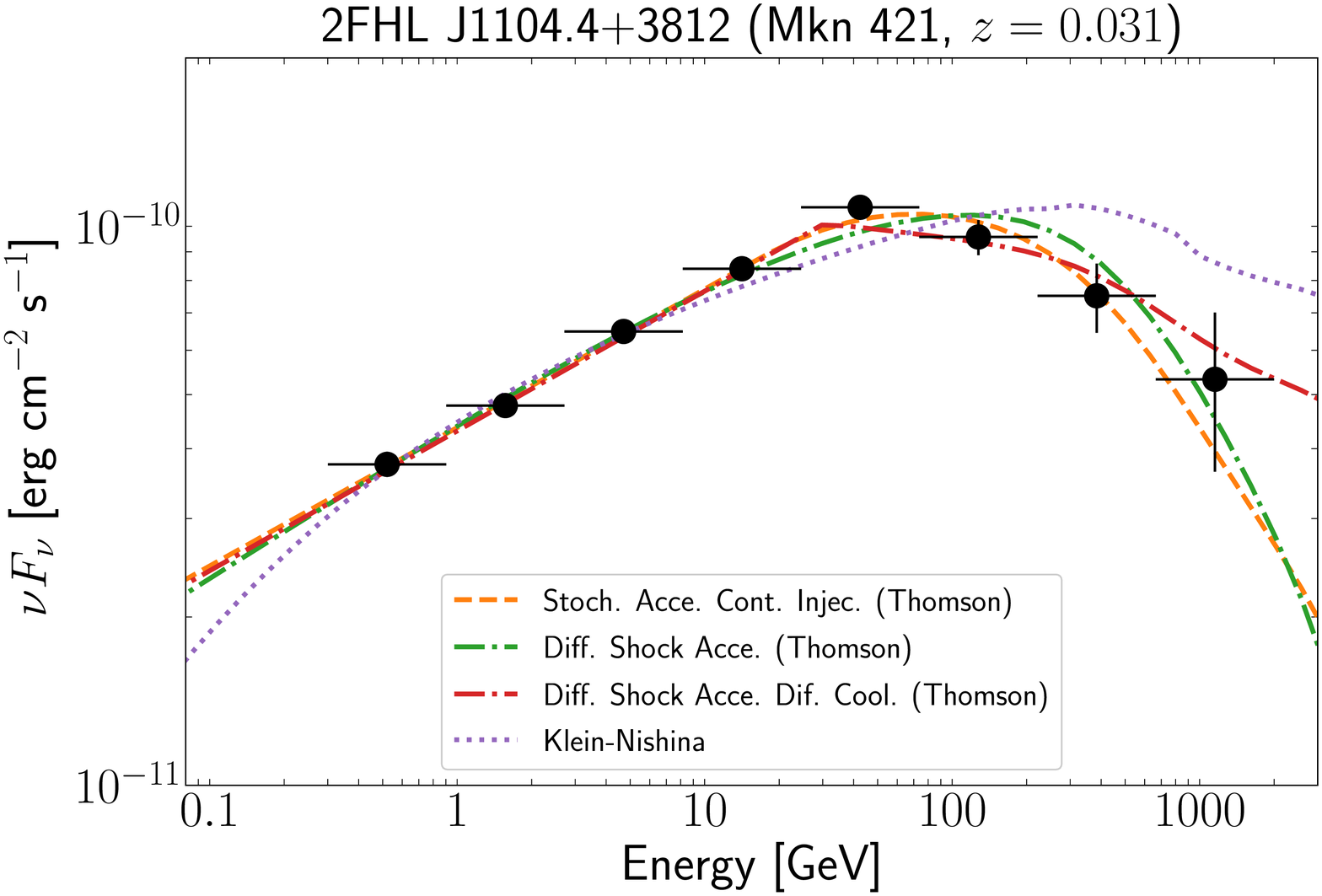} 
    \includegraphics[width=8.2cm]{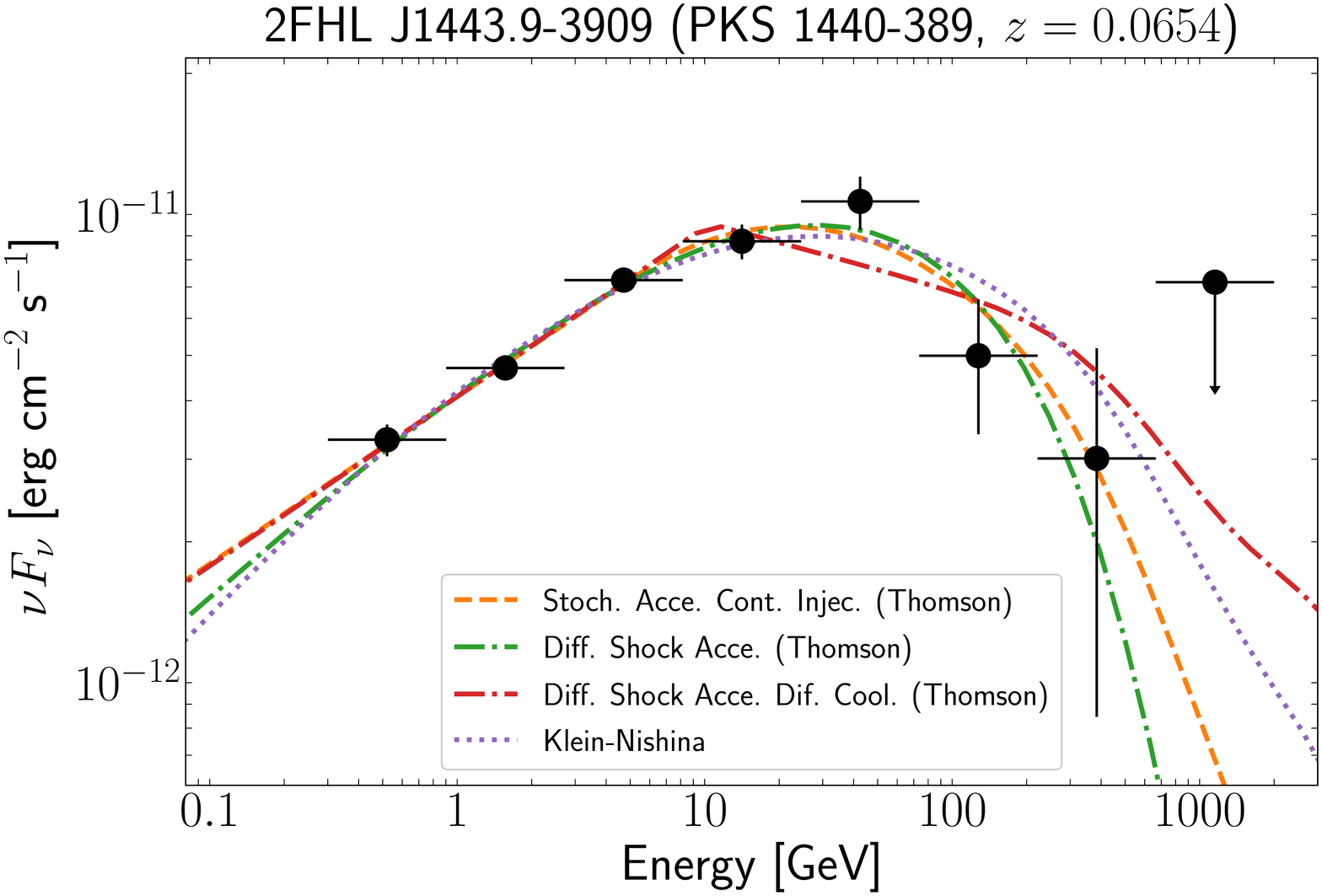}
\caption{\label{fig:SampleSEDs} Examples of high-energy SEDs of four sources in our sample. The LAT data (black circles) 
	are fitted to four emission models: stochastic acceleration with continuous injection in the Thomson 
	regime (log-parabola with low-energy power law, dashed-orange line), radiation-reaction-limited first-order Fermi acceleration 
	in the Thomson regime (power law with exponential cut-off, dashed-dotted green line), radiation-reaction-limited first-order 
	Fermi acceleration with different cooling processes in the Thomson regime (broken power law, dashed-dotted red line), and first-order 
	Fermi acceleration with Compton scattering in the Klein-Nishina regime (power law, dotted magenta line). The EBL attenuation is 
	considered using the model presented by \citet{dominguezetal2011}. Notice that the apparent up-turn in the models at high 
	energies are caused by transforming the models fitted to the intrinsic flux to the observed flux. This is due to the optical 
	depth becoming almost constant at those energies for the given redshifts. (Note that the step like feature of the Klein-Nishina 
    model for Mkn 421 is due to the numerical evaluation of the integral.)}
\end{figure*}

\section{Leptonic Models of $\gamma$-ray Emission}
\label{sec:Model}

In this Section, we describe and derive our physically motivated models for the $\gamma$-ray emission in jets.
In this study, we only consider leptonic emission processes, in which $\gamma$-rays are produced by Compton
scattering off relativistic electrons. The recent possible association of the blazar TXS~0506+056 with the 
track-like EHE neutrino event IceCube-170922A \citep{Aartsen18a} as well as a possible neutrino flare in 
2014 -- 2015 from the same source \citep{Aartsen18b}, suggest that at least in some blazars, hadronic 
emission processes play a role. These could lead to more complicated spectral features than considered 
here, due to the multi-component nature of the $\gamma$-ray emission (proton synchrotron + secondary-electron 
synchrotron from cascades + muon synchrotron + pion synchrotron), and their study is beyond the scope of this 
paper.

\subsection{Theoretical Background}
\label{subsec:Background}

{The first model is based on an electron distribution given by a power law with exponential cut-off, consistent 
with radiation-reaction-limited first-order Fermi acceleration (e.g., DSA). The second model is based on a 
log-parabolic electron distribution, which would be indicative of stochastic acceleration of electrons in the jet 
as suggested by \citet{massaroetal2004, massaroetal2006} and \citet{tramacereetal2007, tramacereetal2011}. The 
fourth model uses an electron distribution described by a broken power law, which could result from different 
acceleration / cooling mechanisms dominating in different energy ranges. The resulting Compton $\gamma$-ray spectra of 
these models are derived in the Thomson regime, so that the $\gamma$-ray spectrum directly reflects the underlying 
electron distribution. The last model assumes a simple power law electron distribution with the main spectral features 
caused by the decrease of the Compton cross section in the high-energy Klein-Nishina regime. It is well established 
that Compton-scattering scenarios are generally well suited to reproduce the $\gamma$-ray spectra of blazars with 
reasonable physical parameters. We therefore do not evaluate the normalization of the resulting Compton spectra in 
detail, as we do not attempt to constrain specific parameters of the physical setup with our fits, but merely 
investigate the spectral shape.

For given electron and synchrotron photon distributions, $n_e (\gamma_e; \Omega_e)$ and $n_{\rm ph} (\epsilon_{\rm ph}; 
\Omega_{\rm ph})$ respectively, the observed Compton flux $\nu F(\nu) = \epsilon F_{\epsilon}$ as a function of the 
up-scattered photons' dimensionless photon energy $\epsilon = h \nu / (m_e c^2)$, is given in terms of the Compton 
cross section. In the Thomson regime, the differential Compton cross section can be approximated by a delta function 
\citep[see][]{bottcheretal2012, dermermenon2012}, where a target photon of dimensionless energy $\epsilon_{\rm ph}$
is up-scattered by an electron with Lorentz factor $\gamma_e = (1 - \beta_e^2)^{-1/2}$, interacting at an angle 
$\mu = \cos \theta$, to a scattered photon energy of $\epsilon_{\rm sc} = \gamma^2 \, (1 - \beta_e \mu) \, 
\epsilon_{\rm ph}$. It is assumed for simplicity that the electron and target photon distributions are isotropic. 
As the shape of the scattered photon spectrum is dominated by the shape of the electron spectrum, one can approximate 
any narrow target photon distribution (such as, e.g., the BLR or dust-torus infrared radiation) as mono-energetic, so 
that $n_{\rm ph}(\epsilon_{\rm ph}) \approx n_{\rm ph;0} \delta (\epsilon_{\rm ph} - \epsilon_0)$. With the additional 
restriction to relativistic electrons, the observed Compton flux is of the form
\begin{equation}
\label{eq:NDotThomson}
\nu F_{\nu} (\epsilon_{\rm sc}) = A \epsilon_{\rm sc}^2 \frac{n_e (\gamma_0)}{\sqrt{\epsilon_{\rm sc} \epsilon_0}} ,
\end{equation}
where $\gamma_0 = \sqrt{\epsilon_{\rm sc} / \epsilon_0}$ and $A$ is a normalization constant. This implies that the 
observed flux will have a form similar to the electron distribution function.

\subsection{First-Order Fermi Acceleration with Thomson Scattering}
\label{subsec:PLEC}

In the case of first-order Fermi acceleration \citep[e.g., DSA;][]{kirkheavens1989, ellisonetal1990, summerlinbaring2012} 
a limiting process, such as radiative cooling and/or a decreasing chance for high-energy particles to cross the shock 
front a large number of times, gives rise to an electron distribution described by a power law with an exponential cut-off
\begin{equation}
\label{eq:PLECElectronDis}
n_e (\gamma_e) = n_{e;0} \gamma_e^{-p} \exp \left( -\frac{\gamma_e}{\gamma_c} \right) ,
\end{equation}
where $p$ is the spectral index and $\gamma_c$ is the cut-off Lorentz factor. Substituting Eq.~\ref{eq:PLECElectronDis} 
into Eq.~\ref{eq:NDotThomson} and absorbing all constants into a proportionality constant $C_1$, the observed 
flux will have the form
\begin{equation}
\label{eq:PLECFlux}
\nu F_{\nu} = C_1 \nu^{-\alpha + 1} \nu_0^{\alpha} \exp \left( -\sqrt{\frac{\nu}{\nu_c}} \right) ,
\end{equation}
where $\alpha = (p-1)/2$ and $\nu_c = \gamma_c^2 \nu_0$ is the cut-off frequency. This will be referred to as the 
power law with exponential cut-off (PL+EC) model. In practise, $C_2$ and $\nu_0$ cannot be constrained independently
from a fit with this model, as they can be absorbed into a combined normalization constant $C_2' = C_2 \, \nu_0^{\alpha}$, 
and a given fit value of $\nu_c$ (within the Fermi range) can always be achieved for an appropriate combination of 
$\nu_0$ and $\gamma_c$, allowing for Compton scattering in the Thomson regime.

\subsection{Stochastic Acceleration with Continuous Injection and Thomson Scattering}
\label{subsec:LPPL}

\citet{tramacereetal2011} showed, using both a statistical and a diffusion equation approach, that stochastic 
acceleration gives rise to an electron distribution with a log-parabolic form, and by solving a diffusion equation 
with radiative losses, that the electron distribution resulting from stochastic acceleration with continuous injection 
could develop a low-energy power law tail while retaining a high-energy log-parabolic peak. Such an electron distribution 
could also result from a stochastic acceleration rate which is constant at low-energies, but becomes energy dependent at 
higher energies \citep{massaroetal2006}. Analytically, we have
\begin{equation}
\label{eq:LPPLElectronDis}
n_e (\gamma_e) = n_{e;0} \left\lbrace \begin{array}{lcl}
(\gamma_e / \gamma_b)^{-a} & \mathrm{if} & \gamma_e \le \gamma_b \\
(\gamma_e / \gamma_b)^{-[a + b \ln (\gamma_e / \gamma_b)]} & \mathrm{if} & \gamma_e > \gamma_b
\end{array} \right. ,
\end{equation}
where $a$ is the low-energy limit of the slope, $b$ parameterizes the curvature of the distribution, and $\gamma_b$ is 
the Lorentz factor at which the break/transition occurs \citep[see also][]{massaroetal2004, tramacereetal2007}. The 
curvature of the distribution is inversely proportional to both the number of acceleration steps or the acceleration time, 
and the variance in the energy gained during each acceleration step or the momentum diffusion coefficient. In the absence 
of radiative cooling, the distribution will therefore become a power law for very long acceleration times or effective 
momentum diffusion, but if radiative cooling is taken into account, the curvature will increase. Substituting 
Eq.~\ref{eq:LPPLElectronDis} into Eq.~\ref{eq:NDotThomson}, yields
\begin{equation}
\label{eq:LPPLFlux}
\nu F_{\nu} = C_2 \nu \sqrt{\frac{\nu}{\nu_0}} \left\lbrace \begin{array}{lcl}
(\nu / \nu_b)^{-a/2} & \mathrm{if} & \nu \le \nu_b \\
(\nu / \nu_b)^{-[a + b \ln (\nu / \nu_b)/2]/2} & \mathrm{if} & \nu > \nu_b
\end{array} \right. 
\end{equation}
for the observed flux, where all constants were absorbed into the proportionality constant $C_2$ and 
$\nu_b = \gamma_b^2 \nu_0$ is the frequency at which the break/transition occurs. Notice that there is a similar 
dependence between $\nu_b$ and $\nu_0$ as there is between $\nu_c$ and $\nu_0$ in the PL+EC model. This model 
will be referred to as the log-parabola with low-energy power law (LP+PL) model.

\subsection{First-Order Fermi Acceleration with Different Acceleration / Cooling Regimes and Thomson Scattering}
\label{subsec:BPL}

If two different physical processes dominate in different energy ranges, such as radiative vs. adiabatic cooling, 
then the electron distribution can be described by a broken power-law: 
\begin{equation}
\label{eq:BPLElectronDis}
n_e (\gamma_e) = n_{e;0} \left\lbrace \begin{array}{lcl}
(\gamma_e / \gamma_b)^{-q} & \mathrm{if} & \gamma_e \le \gamma_b \\
(\gamma_e / \gamma_b)^{-s} & \mathrm{if} & \gamma_e > \gamma_b
\end{array} \right. ,
\end{equation}
where $q$ and $s$ is the spectral index of the low- and high-energy power law, respectively. Substituting this 
into Eq.~\ref{eq:NDotThomson} and absorbing all constants into a single proportionality constant, results in an 
observed flux with the form
\begin{equation}
\label{eq:LPBPLFlux}
\nu F_{\nu} = C_3 \nu \sqrt{\frac{\nu}{\nu_0}} \left\lbrace \begin{array}{lcl}
(\nu / \nu_b)^{-q/2} & \mathrm{if} & \nu_e \le \nu_b \\
(\nu / \nu_b)^{-s/2} & \mathrm{if} & \nu_e > \nu_b
\end{array} \right. .
\end{equation}
In this model, which will be referred to as the broken power-law (BPL) model, there is the same dependence 
between $\nu_b$ and $\nu_0$ as in the case of the LP+PL model.

\subsection{First-Order Fermi Acceleration with Klein-Nishina Scattering}
\label{subsec:KN}

In the Klein-Nishina regime, the Compton cross section is more complicated, but for scattering by ultra-relativistic 
electrons, it can be well represented by the head-on approximation, in which the scattered photon propagates in the 
direction of the in-coming electron \citep[see][]{bottcheretal2012, dermermenon2012}. Using the same setup as in the 
Thomson regime, the angle integrations can be done analytically to give the observed flux as a function of the 
normalized up-scattered photon energy, as done by \citet{jones1968} \citep[see also][]{bottcheretal2012, dermermenon2012}. 
The decrease of the Compton cross section in the Klein-Nishina regime will lead to high-energy spectral curvature in 
the Compton spectrum even for an electron distribution described by a simple power law,
\begin{equation}
\label{eq:PLElectronDis}
n_e (\gamma_e) = n_{e;0} \gamma_e^{-p}.
\end{equation}
Again assuming a mono-energetic target-photon distribution and absorbing constant factors into a proportionality 
constant $C_4$, the observed Compton flux can be written as
\begin{equation}
\label{eq:PLFlux}
\nu F_{\nu} = C_4 \frac{\epsilon^2}{\epsilon_0} \int_{\gamma_1}^{\infty} \gamma_e^{-(p+2)} \times
\end{equation}
$$
\left[ 2 q \ln (q) + (1 + 2 q)(1 - q) + \frac{(1 - q)(4 \epsilon_0 \gamma_e q)^2}{2 (1 + 4 \epsilon_0 \gamma_e q)} 
\right] d \gamma_e ,
$$
where
\begin{equation}
\label{eq:Q}
q = \frac{\epsilon}{4 \epsilon_0 \gamma_e (\gamma_e - \epsilon)} ,
\end{equation}
the subscript were dropped and it should be noted that only the up-scattering part of the integral of \citet{jones1968} 
relevant to the $\gamma$-ray regime is used. The limits on $\epsilon$ where the integral is non-zero, impose a lower 
limit on the $\gamma_e$ integration, given by
\begin{equation}
\label{eq:LowerGamma}
\gamma_1 = \frac{\epsilon + \sqrt{\epsilon^2 + \epsilon / \epsilon_0}}{2}.
\end{equation}
This will be referred to as the Klein-Nishina (KN) model.

\subsection{Influence of Relativistic Doppler Boosting}
\label{subsec:Doppler}

In the case of external Compton (EC) scattering, relativistic Doppler boosting of the external photon fields 
into the rest frame of the 
emission region and back into the observer's frame needs to be considered. If the emission region moves with a bulk 
Lorentz factor $\Gamma = (1 - \beta_{\Gamma}^2)^{-1/2}$ (typically with a value of $\Gamma \sim 10$) at an angle $\mu = 
\cos \theta_{\rm obs}$ with respect to our line of sight, Doppler boosting is characterized by the Doppler factor 
$\delta_D = [ \Gamma \, (1 - \beta_{\Gamma}\mu) ]^{-1}$. Blazars are observed at a small angle $\theta_{\rm obs} \sim 1 
/ \Gamma$ with respect to the jet axis, so that the Doppler factor is typically of the order of the Lorentz factor, 
$\delta_D \approx \Gamma$. External target photons of energy $\epsilon_0$ are then Doppler boosted into the emission 
region rest frame (denoted here with primed quantities) as $\epsilon'_0 \approx \Gamma \epsilon_0$ and back into the 
observer's frame as $\epsilon_{\rm sc}^{\rm obs} \approx \Gamma \epsilon'_{\rm sc}$. In the Thomson regime, we 
therefore have $\epsilon_{\rm sc}^{\rm obs} \approx \epsilon_0 \, \Gamma^2 \, (\gamma'_e)^2$. Interpreting the values 
of $\epsilon_0$ resulting from the formalism developed above (neglecting Doppler boosting) and the fitting routine 
described below, as the actual value of the target photon energy (in the AGN rest frame), the values of $\gamma_e$ 
found in the fitting routine then correspond to co-moving electron energies of $\gamma'_e = \gamma_e/\Gamma$. 

In the case of SSC, both the synchrotron target photons and the SSC-scattered high-energy photons are subject to the 
same Doppler boost $\delta_D \sim \Gamma$. Hence, the fit value of $\nu_0$ and $\epsilon_0$ may be 
interpreted as the observed synchrotron photon frequency and energy, respectively, and the Thomson limit applies when 
$\epsilon_0 \gamma_e / \Gamma \ll 1$. 

\subsection{Extragalactic Background Light Attenuation}
\label{subsec:EBL}

High-energy photons travelling over cosmological distances are attenuated by pair-production interactions with the 
extragalactic background light \citep[EBL,][]{nikishov1962, gouldschreder1966}. The EBL is the diffuse infrared through 
ultraviolet radiation accumulated over the history of the Universe \citep[e.g.,~][]{hauserdwek2001}. The intrinsic flux 
is related to the observed flux by
\begin{equation}
\left. \frac{dF}{dE} \right|_{\rm int} = \left. \frac{dF}{dE} \right|_{\rm obs} \exp \left[ \tau (E,z) \right] .
\end{equation}
We consider this effect using the optical depths $\tau$ by \citet{dominguezetal2011}, which are provided as a function 
of observed $\gamma$-ray energy $E$ and redshift $z$ of the source.

\section{Fitting Methodology}
\label{sec:Fitting}

The four models are fitted to the data using a $\chi^{2}$ minimization fitting routine. Upper limits ($1\sigma$)  
are also considered in the fitting by using half of the limit as both the flux data point and the flux error. This is a 
possible way to handle upper limits and thus, use as much spectral information as possible. Considering the large errors 
on flux points implied by the upper limits, the fitting routine will assign small weights to these data points. Therefore, 
we do not expect that the results would change qualitatively if the upper limits are treated differently.

In order to choose the best-fitting model we apply a maximum likelihood ratio test with a $95 \%$ confidence level. 
We use the logarithm of the likelihood ratio $t = -0.5(\chi_1^2 - \chi_0^2)$ as the test statistic, where $\chi_0^2$ and 
$\chi_1^2$ are the chi-square values of the null and alternative models, respectively. The null model is the model with 
the larger number of degrees of freedom (DoF). Notice that the PL+EC, LP+PL, BPL, and KN models have 4, 5, 5, and 3 free 
parameters, respectively, which means that with 8 flux data points, the models have 4, 3, 3, and 5 DoF, respectively. If 
the likelihood ratio is too large, the null model is rejected and the alternative model accepted, otherwise the null 
model is accepted and the alternative model rejected. Since $\Delta \chi^2$ is approximately $\chi^2$-distributed, with 
a DoF equal to the difference in the DoF of the two models being compared, a $95 \%$ confidence level is equivalent to 
$t > 0.5 \times 3.84 = 1.92$ for 1 DoF and $t > 0.5 \times 5.99 = 2.995$ for 2 DoF. Notice that the LP+PL and BPL models 
have the same number of DoF so that the likelihood ratio test cannot be performed on these two models. These two models 
are compared according to their $\chi^{2}$-values and the model with the smallest $\chi^{2}$-value is accepted as the 
favoured model. This test is done for each blazar between all the different combinations of models. The model that was
preferred when compared to all other models is then accepted as the favoured model. Obviously, for any individual blazar, 
this cannot be considered a statistically robust statement of preference for a certain model. However, a systematic 
preference of one model throughout the sample of 128 blazars that we have investigated here, would provide a clear 
indication concerning the true spectral shape.

\begin{deluxetable*}{cCCCCCCC}
\tablecaption{\label{tab:Reject}Rejection criteria applied to fits and the number of fits rejected. 
See text for motivation of these rejection criteria.}
\tablehead{
\colhead{Model} & \colhead{Rejection Criteria} & \colhead{All (128)} & \colhead{Variable (47)} & 
\colhead{Non-variable (81)} & \colhead{BL Lacs (106)} & \colhead{FSRQs (10)} & \colhead{Other (12)}}
\decimals
\startdata
\multirow{3}{*}{PL+EC} & Q < 0.001                  & 2  & 2 & 0  & 2  & 0 & 0 \\
                       & \alpha < 0.5               & 19 & 1 & 18 & 18 & 0 & 1 \\
                       & \mathrm{Total \; rejected} & 21 & 3 & 18 & 20 & 0 & 1 \\
\hline
\multirow{5}{*}{LP+PL} & Q < 0.001                  & 1  & 1 & 0  & 0  & 1 & 0 \\
                       & a < 1                      & 1  & 0 & 1  & 1  & 0 & 0 \\
                       & b < 0                      & 34 & 6 & 28 & 26 & 1 & 7 \\
                       & \mathrm{Total \; rejected} & 36 & 7 & 29 & 27 & 2 & 7 \\
\hline
\multirow{4}{*}{BPL} & Q < 0.001                  & 3  & 2  & 1  & 1  & 1 & 1 \\
                     & q < 2                      & 21 & 2  & 19 & 19 & 0 & 2 \\
                     & s < q                      & 30 & 6  & 24 & 25 & 1 & 4 \\
                     & \mathrm{Total \; rejected} & 54 & 10 & 44 & 45 & 2 & 7 \\
\hline
\multirow{4}{*}{KN} & Q < 0.001                     & 6  & 6  & 0  & 3  & 2  & 1 \\
                    & p < 1                         & 12 & 2  & 10 & 10 & 0  & 2 \\
                    & \epsilon_0 > 1 \times 10^{-4} & 69 & 30 & 39 & 50 & 10 & 9 \\
                    & \mathrm{Total \; rejected}    & 71 & 32 & 39 & 52 & 10 & 9 \\
\enddata
\end{deluxetable*}

\section{Results}
\label{sec:Results}

In this Section, the results of the fits of the four physically motivated models to the 2FHL blazar data are 
presented.

\begin{splitdeluxetable*}{cCR@{}C@{}LR@{}C@{}LR@{}C@{}LR@{}C@{}LR@{}C@{}LR@{}C@{}LBR@{}C@{}LR@{}C@{}LR@{}C@{}LR@{}C@{}LR@{}C@{}LR@{}C@{}L}
\tablecaption{\label{tab:ParSum}Averages of the model parameters of the fits, as well as the range of $\chi_r^2$-values for comparison. See the text for details.}
\tablehead{
\multirow{2}{*}{Model} & \multirow{2}{*}{Parameter} & \multicolumn{18}{c}{All Fits} & \multicolumn{18}{c}{Only Accepted Fits} \\
\cline{3-38}
\colhead{} & \colhead{} & \multicolumn{3}{c}{All (128)} & \multicolumn{3}{c}{Variable (47)} & \multicolumn{3}{c}{Non-variable (81)}  & \multicolumn{3}{c}{BL Lacs (106)} & \multicolumn{3}{c}{FSRQs (10)} & \multicolumn{3}{c}{Other (12)} & \multicolumn{3}{c}{All} & \multicolumn{3}{c}{Variable} & \multicolumn{3}{c}{Non-variable} & \multicolumn{3}{c}{BL Lacs} & \multicolumn{3}{c}{FSRQs} & \multicolumn{3}{c}{Other}}
\startdata
\multirow{3}{*}{PL+EC} & \chi_r^2 & 0.3 & - & 5 & 0.3 & - & 5 & 0.3 & - & 4 & 0.3 & - & 5 & 0.8 & - & 4 & 0.5 & - & 4 
                                  & 0.3 & - & 4 & 0.3 & - & 4 & 0.3 & - & 4 & 0.3 & - & 4 & 0.8 & - & 4 & 0.5 & - & 4 \\
 & \alpha & 0.8 & \pm & 0.3 & 0.9 & \pm & 0.2 & 0.7 & \pm & 0.3 & 0.7 & \pm & 0.2 & 1.1 & \pm & 0.2 & 0.9 & \pm & 0.3 
          & 0.9 & \pm & 0.2 & 0.9 & \pm & 0.2 & 0.8 & \pm & 0.2 & 0.8 & \pm & 0.2 & 1.1 & \pm & 0.2 & 1.0 & \pm & 0.3 \\
 & \nu_c\tablenotemark{a}\tablenotemark{b} & 26 & \pm & 1 & 26 & \pm & 1 & 26 & \pm & 2 & 26 & \pm & 1 & 25 & \pm & 2 & 26 & \pm & 1 
                                           & 26 & \pm & 1 & 25 & \pm & 1 & 26 & \pm & 2 & 26 & \pm & 1 & 25 & \pm & 2 & 26 & \pm & 2 \\
\hline
\multirow{4}{*}{LP+PL} & \chi_r^2 & 0.2 & - & 6 & 0.2 & - & 6 & 0.2 & - & 5 & 0.2 & - & 5 & 0.8 & - & 6 & 0.3 & - & 2 
                                  & 0.2 & - & 4 & 0.2 & - & 4 & 0.3 & - & 3 & 0.2 & - & 4 & 0.8 & - & 2 & 0.5 & - & 2 \\
 & a & 2.7 & \pm & 0.6 & 2.9 & \pm & 0.5 & 2.6 & \pm & 0.7 & 2.6 & \pm & 0.5 & 3.5 & \pm & 0.4 & 3.2 & \pm & 0.6 
     & 2.6 & \pm & 0.5 & 2.9 & \pm & 0.4 & 2.4 & \pm & 0.6 & 2.5 & \pm & 0.5 & 3.4 & \pm & 0.4 & 3.0 & \pm & 0.2 \\
 & b & 0.3 & \pm & 0.8 & 0.3 & \pm & 0.3 & 0.3 & \pm & 0.9 & 0.3 & \pm & 0.8 & 0.3 & \pm & 0.3 & -0.1 & \pm & 0.4 
     & 0.5 & \pm & 0.7 & 0.3 & \pm & 0.2 & 0.6 & \pm & 0.9 & 0.5 & \pm & 0.8 & 0.3 & \pm & 0.2 & 0.2  & \pm & 0.1 \\
 & \nu_b\tablenotemark{a}\tablenotemark{b} & 23.9 & \pm & 0.5 & 23.9 & \pm & 0.4 & 23.9 & \pm & 0.5 & 23.9 & \pm & 0.5 & 23.7 & \pm & 0.4 & 23.7 & \pm & 0.6
                                           & 23.9 & \pm & 0.5 & 23.9 & \pm & 0.4 & 23.9 & \pm & 0.5 & 24.0 & \pm & 0.5 & 23.6 & \pm & 0.2 & 23.9 & \pm & 0.2 \\
\hline
\multirow{4}{*}{BPL} & \chi_r^2 & 0.3 & - & 8 & 0.3 & - & 8 & 0.3 & - & 6 & 0.3 & - & 6 & 1 & - & 7 & 0.3 & - & 8 
                                & 0.3 & - & 5 & 0.3 & - & 5 & 0.4 & - & 4 & 0.3 & - & 4 & 1 & - & 5 & 0.5 & - & 2 \\
 & q & 2.7 & \pm & 0.9 & 2.9 & \pm & 0.5 & 3   & \pm & 1   & 2.5 & \pm & 0.9 & 3.5 & \pm & 0.4 & 3.1 & \pm & 0.7 
     & 2.8 & \pm & 0.5 & 2.9 & \pm & 0.5 & 2.6 & \pm & 0.4 & 2.7 & \pm & 0.4 & 3.5 & \pm & 0.4 & 3.1 & \pm & 0.3 \\
 & s & 3.1 & \pm & 0.8 & 3.4 & \pm & 0.7 & 2.9 & \pm & 0.8 & 3.0 & \pm & 0.8 & 4.1 & \pm & 0.8 & 3.1 & \pm & 0.7 
     & 3.4 & \pm & 0.8 & 3.5 & \pm & 0.6 & 3.4 & \pm & 0.9 & 3.3 & \pm & 0.8 & 4.1 & \pm & 0.7 & 3.5 & \pm & 0.4 \\
 & \nu_b\tablenotemark{a}\tablenotemark{b} & 24.0 & \pm & 0.5 & 24.0 & \pm & 0.5 & 24.0 & \pm & 0.4 & 24.0 & \pm & 0.4 & 24.1 & \pm & 0.1 & 23.8 & \pm & 0.8 
                                           & 24.1 & \pm & 0.4 & 24.1 & \pm & 0.3 & 24.1 & \pm & 0.4 & 24.1 & \pm & 0.4 & 24.0 & \pm & 0.1 & 24.1 & \pm & 0.5 \\
\hline
\multirow{3}{*}{KN} & \chi_r^2 & 0.3 & - & 24 & 0.5 & - & 24 & 0.3 & - & 4 & 0.3 & - & 12 & 0.7 & - & 24                 & 0.4 & - & 5 
                               & 0.4 & - & 4  & 0.5 & - & 3  & 0.4 & - & 4 & 0.4 & - & 4  & -   & - & -\tablenotemark{c} & 0.5 & - & 2 \\
 & p & 1.7 & \pm & 0.5 & 1.7 & \pm & 0.5 & 1.7 & \pm & 0.5 & 1.7 & \pm & 0.5 & 1.7 & \pm & 0.3                & 1.6 & \pm & 0.4 
     & 2.1 & \pm & 0.3 & 2.2 & \pm & 0.3 & 2.0 & \pm & 0.3 & 2.1 & \pm & 0.3 & -   & \pm & -\tablenotemark{c} & 1.9 & \pm & 0.3 \\
 & \epsilon_0\tablenotemark{a} & -3   & \pm & 2   & -3   & \pm & 2   & -4   & \pm & 2   & -4   & \pm & 2   & -2 & \pm & 1                  & -3   & \pm & 2 
                               & -5.3 & \pm & 0.6 & -5.2 & \pm & 0.7 & -5.3 & \pm & 0.6 & -5.3 & \pm & 0.6 & -  & \pm & -\tablenotemark{c} & -5.5 & \pm & 0.7 \\
\enddata
\tablenotetext{a}{The averages and standard deviations of parameters with exponential values are given for the logarithm of the parameters.}
\tablenotetext{b}{Hz}
\tablenotetext{c}{Only one or no accepted model.}
\end{splitdeluxetable*}

\subsection{Accepted models}

In some cases, even though formally an acceptable spectral fit could be achieved with a given model, the 
best-fit parameters are problematic and/or unphysical. We expect $\gamma$-ray emission produced by radiatively cooled 
electrons. Thus, a radiation spectrum indicating an electron spectrum harder than $\gamma^{-2}$ would have to be 
accelerated/injected from a population following a spectrum harder than $\gamma^{-1}$, which is difficult to reconcile 
with any known particle acceleration mechanism. It is highly unlikely that the spectra of the PL+EC, LP+PL, BPL, 
and KN models have a photon spectral index $\alpha$ harder than $0.5$, or a spectral index of the radiating particle
distribution of $a < 1$, $q < 2$, or $p < 1$, respectively. We point out that we have assumed that the electron
spectra do not have low-energy cut-offs, i.e., our electron spectra always start at $\gamma_{\rm min} = 1$. 
While a large value of $\gamma_{\rm min} \gg 1$ could, in principle, also produce very hard low-energy $\gamma$-ray
spectra \citep[see, e.g.,][]{Katarzynski12}, there is no accepted scenario which would realistically produce such 
a large low-energy cut-off. We have therefore not considered this possibility in this study. 

There are a few cases for the LP+PL model where $b$ have 
negative values, causing the fitted spectra to curve upwards, which is not physical. There is a large number of 
fitted values for $\epsilon_0$ that fall in the Thomson regime. In the other models, $\nu_0$ 
enters only as an arbitrary normalization constant due to the dependence of $\gamma_c$ or $\gamma_b$ on $\nu_0$, which 
could be absorbed into $C$; this was not done in order to eliminate a dependence of $C_1$ on $\alpha$ in the case of 
the PL+EC model or a dependence between $C$ and $\nu_b$ in the case of the LP+PL and BPL models. The fitted value 
might therefore be considered unphysical if $\epsilon_0 > 1 \times 10^{-4}$ for the KN model. 
For the BPL model we additionally require that the high-energy component is softer than the low-energy 
component and hence $s < q$ must hold. The probability for the $\chi^2$-value to be larger than a 
certain $\chi^2$-value by chance, $Q$, was also calculated for each fit. If $Q \gtrsim 0.1$, the fit is believable; if 
$0.1 \gtrsim Q \gtrsim 0.001$, the fit may be acceptable if the uncertainties are not normally distributed or have been 
moderately underestimated; if $Q \lesssim 0.001$, then the fit can be statistically rejected; if $Q$ is very close to 
$1$, the fit might be too good to be true and this can be caused by an overestimation of the uncertainties or fraud in 
the data points. However, since the proportionality constants $C_i$ are arbitrary, the latter case can only be 
interpreted as a good fit.

Based on these criteria some of the fits were rejected, as summarized in Table~\ref{tab:Reject}. The average with 
standard deviation for the fitted parameters of the four models to the SEDs are summarized in column three to 
eight of Table~\ref{tab:ParSum} and the summary of the fitted parameters of only the accepted fits are given in the last 
six columns of Table~\ref{tab:ParSum}. The range of reduced chi-squared values $\chi_r^2$ are also included in the 
table for comparison while the normalization constants and $\nu_0$ of the PL+EC, LP+PL, and BPL models are not 
shown since they are arbitrary and not of interest. The average and standard deviation of parameters with exponential 
values ($\nu_c$, $\nu_b$ and $\epsilon_0$) are given as the average and standard deviation 
of the base 10 logarithm of the parameters.

The restrictions that the high-energy component of the BPL model is softer than the low-energy component and that 
Compton scattering occur indeed in the Klein-Nishina regime in the KN model, have led to the rejection of a lot of the fits 
of these models. The frequencies corresponding to $\epsilon_0$ for the accepted KN fits are of the order of $\sim 10^{16}$ 
-- $10^{17}$~Hz and fall in the ultraviolet to soft X-ray range, characteristic of synchrotron photons in the case of ISP or 
HSP blazars. The validity of the PL+EC model with Compton scattering in the Thomson regime up to $\sim 1$~TeV implies that the 
target photons must have frequencies $\nu_0 \lesssim 10^{14}$~Hz, favouring the dust-torus emission as their source. In 
several fits of the PL+EC model, very large values of the cut-off frequency $\nu_c$ (up to $\sim 10^{31}$ Hz, compared to the 
data ranging up to $\sim 3 \times 10^{27}$ Hz) resulted, indicating that the fit could be well approximated by a pure power law. 
This is also seen in the LP+PL model as small curvature parameter $b$ values.

\begin{splitdeluxetable*}{cCCCCCCBCCCCCC}
\tablecaption{\label{tab:BestFit}Number of times each model fitted the best.}
\tablehead{
\multirow{2}{*}{Model} & \multicolumn{6}{c}{All Fits} & \multicolumn{6}{c}{Only Accepted Fits} \\
\cline{2-13}
\colhead{} & \colhead{All (128)} & \colhead{Variable (47)} & \colhead{Non-variable (81)} & \colhead{BL Lacs (106)} & \colhead{FSRQs (10)} & \colhead{Other (12)} & \colhead{All (128)} & \colhead{Variable (47)} & \colhead{Non-variable (81)} & \colhead{BL Lacs (106)} & \colhead{FSRQs (10)} & \colhead{Other (12)}} 
\decimals
\startdata
PL+EC & 24 & 15 & 9  & 19 & 3 & 2
      & 64 & 25 & 39 & 48 & 6 & 10 \\
LP+PL & 8  & 5  & 3  & 5  & 2 & 1
      & 13 & 6  & 7  & 11 & 2 & 0 \\
BPL   & 22 & 6  & 16 & 21 & 0 & 1
      & 15 & 8  & 7  & 13 & 2 & 0 \\
KN    & 73 & 21 & 52 & 60 & 5 & 8
      & 35 & 7  & 28 & 33 & 0 & 2 \\
\hline
No accepted model & 1 & 0 & 1 & 1 & 0 & 0 
                  & 1 & 1 & 0 & 1 & 0 & 0 \\
\enddata
\end{splitdeluxetable*}

\subsection{Variable and Non-variable Blazars}
\label{subsec:VarNonVarFits}

The physical processes underlying the four models are quite different and most likely time-dependent. It might be 
expected that stochastic acceleration would always be present if there is turbulence in the jet and the decrease of the 
Compton cross section in the Klein-Nishina regime will be relevant whenever the target photon energy is $\epsilon_0 
\gtrsim 10^{-5}$ in which case $\gamma$-ray photons of $> 10$~GeV can no longer be produced by Thomson scattering. 
However, the relevant acceleration and cooling processes are highly time-dependent and a combination of all of these 
processes could lead to artificial spectral features in the time-averaged spectra which we are fitting. In an attempt 
to avoid such complications, the blazars were divided into variable and non-variable blazars. Unfortunately, 2FHL 
presents a variability analysis only considering photons above 50~GeV, not for our broader energy range data 
(300~MeV--2~TeV). Developing a complete time analysis for the data is beyond the scope of this paper. However, we can 
work around this limitation by using the 3FHL variability study \citep{ajello2017} since the 3FHL contains 127 of the 
128 2FHL sources (the only drawback is that photons below 10~GeV are not considered in the variability analysis). 
According to the 3FHL catalog there are 47 variable blazars in our sample and we assume that the other 81 blazars are 
non-variable or nearly so.

The average with standard deviation for fitted parameters of the four models of the variable and non-variable blazars 
are summarized in columns 4 and 5 of Table~\ref{tab:ParSum}, respectively. Also shown in columns 6, 7, and 8, is a 
summary of the fitted parameters for BL Lacs, FSRQs, and other blazar types, respectively. The numbers of fits rejected 
by the various rejection criteria are also summarized in Table~\ref{tab:Reject}. Lower spectral indices (harder spectra) 
are needed to fit non-variable blazars than variable ones and BL Lacs also require lower spectral indices than other blazar 
classes. Essentially, most of the fits rejected due to unphysically hard spectra are those of non-variable and BL Lac 
blazars. These trends also appear when comparing the averages of the fitted parameters and qualitatively, the average values 
of the fitted parameters do not differ much between the subgroups of the accepted fits.
The KN model seems incapable of reproducing the spectra of FSRQs and blazars of other types.

\subsection{Preferred Model}
\label{subsec:BestFit}

Qualitatively, when comparing the four models in Fig.~\ref{fig:SampleSEDs} and the $\chi_r^2$-values in 
Table~\ref{tab:ParSum}, it seems that all four models fit the SEDs similarly well. The number of times each model 
provided the best fit, based on the likelihood ratio test outlined in Section~\ref{sec:Fitting} and where a model was 
counted as being a good fit if the other three models were rejected, should quantitatively indicate which model may 
be considered systematically preferred. These results are summarized in Table~\ref{tab:BestFit}.

Focussing only on the accepted fits, the LP+PL and BPL models seem to be systematically disfavoured for most blazars. 
This indicates strong evidence against Thomson scattering by a log-parabola or broken power law electron distribution. 
The PL+EC model was preferred for the majority of the variable blazars, the FSRQs in the sample, as well as for blazars 
of unknown type (other). For the non-variable blazars as well as for BL Lac type blazars, the PL+EC and KN models were 
preferred approximately equally often, with a slight preference for the PL+EC model.

\section{Summary and Discussion}
\label{sec:Summary}

In this work we analyzed the first nine years of \emph{Fermi}-LAT data in the energy range from 300~MeV to 2~TeV in 
order to extend the energy spectral coverage of the 128 2FHL blazars. These spectral data were compared to four models 
for the production of $\gamma$-ray spectra assuming a single-zone leptonic model: (1) radiation-reaction-limited first-order 
Fermi acceleration of electrons (power law with exponential cut-off) with Compton scattering in the Thomson regime, (2) 
stochastic acceleration of electrons with continuous injection (log-parabola with low-energy power law) and Compton 
scattering in the Thomson regime, (3) first-order Fermi acceleration of electrons with different acceleration/cooling 
mechanisms dominating in different energy regimes (broken power law) and Compton scattering in the Thomson regime, and 
(5) Compton scattering by a pure power law distribution of electrons with spectral curvature due to scattering in the 
Klein-Nishina regime.

Obviously, these are not the only plausible spectral shapes. However, they represent four fundamentally different, 
physically plausible ways of the formation of $\gamma$-ray spectra in blazars, and there is no (finite) exhaustive list 
of all possible combinations of effects that might contribute in reality. The PL+EC, LP+PL, and BPL models, 
corresponds to physically motivated electron distributions (DSA, stochastic acceleration with continuous injection, 
and energy dependent acceleration/cooling, respectively), assuming Compton scattering in the Thomson regime, and the 
pure power law would simply be extreme cases of either model ($b = 0$, $\nu_c \rightarrow \infty$, or 
$\nu_b \rightarrow \infty$, respectively). The power law with Compton scattering in the Klein-Nishina regime was 
introduced to check whether Klein-Nishina effects might, instead, be dominant in the formation of spectral curvature. 
We therefore consider these four shapes ``basic building blocks'' of the spectral shapes of blazars. A (more 
realistic) combination of LP or PL+EC with Klein-Nishina effects would introduce too many free parameters, so that the 
available spectra would not be able to provide a meaningful distinction. A systematic preference for any (one or two) 
of these fundamental models throughout the entire 2FHL sample may be considered a significant indication of the 
dominant mode of $\gamma$-ray spectra formation in these blazars.

The fitted parameters found here only refer to the general shape of the high-energy spectrum and constrain the energy
of target photons for Compton scattering and the energy distribution of the electrons. However, other physical 
parameters (such as, e.g., the magnetic field or the bulk Lorentz and Doppler factor), cannot be meaningfully
constrained based on fits to the $\gamma$-ray spectra alone. This degeneracy can be broken by fitting a broader 
energy range, including the synchrotron component of the SED \citep[see, e.g.,][for an application]{paliyaetal2018}. 
While the shape of the high-energy tail of the synchrotron spectrum can often be probed well in HSP blazars (where 
the synchrotron peak is often prominent in the X-ray regime), this is generally difficult in LSP and ISP blazars 
\citep{abdoetal2010b} (where the synchrotron peak is located in the infrared though optical regime and the high-energy 
tail is often unobservable), as it can be located in the inaccessible ultraviolet regime and/or because it is overwhelmed 
by the low-energy tail of the high-energy spectral component. It is therefore difficult to characterize the full SED. 
Thus, the large sample of well-determined blazar $\gamma$-ray spectra measured by {\it Fermi}-LAT seems to provide the 
best and most abundant test bed for the high-energy shapes of blazar spectra, even though it only allows us to characterize 
the underlying physical processes and not to pin down specific parameter values.

The blazars were divided into a variable and non-variable subgroup, as a combination of the different, time-dependent 
physical processes could lead to artificial spectral features in the time-averaged spectra of variable blazars. The 
blazars were also further divided into BL Lacs, FSRQs, and other types of blazars, as the physical acceleration 
mechanisms could vary among the different types of blazars. Our most significant result is the rejection of the model 
with Thomson scattering by an electron distribution with a broken power law or a log-parabola with a low-energy power law. 
This does not imply a complete rejection of these electron distributions. However, it indicates that, if such an electron 
distribution is present, additional effects, such as the Klein-Nishina cut-off, must play a significant role in the 
formation of blazar $\gamma$-ray spectra.

The first-order Fermi acceleration with Thomson scattering and the decrease of the Compton cross section in the 
Klein-Nishina regime could successfully explain the high-energy spectral shape of almost equal numbers of non-variable blazars 
as well as of BL Lacs. This is consistent with the standard interpretation of SSC-dominated $\gamma$-ray emission in BL Lac 
objects, where a gradual transition from the Thomson to the Klein-Nishina regime is expected throughout the high-energy 
$\gamma$-ray range. We remind the reader that for the PL+EC model, the target photon energy cannot be constrained 
from the spectral fits to $\gamma$-ray spectra alone, as there is a degeneracy between $\nu_0$ and $\gamma_c$ (see Section 
\ref{subsec:PLEC}). Combinations of $\nu_0$ and $\gamma_c$ can therefore always be found that allow for Compton scattering 
in the Thomson regime up to the highest Fermi energies. This requires electron cut-off energies of $\gamma_c \gtrsim 10^5$ 
and soft target photons with frequencies $\nu_0 \lesssim 10^{14}$~Hz, thus strongly favouring a dust-torus origin of the 
target photons \citep[in agreement with the results by][]{costamante2018}.

Although DSA might be expected as a plausible acceleration mechanism for variable blazars, there might be a 
contribution of various physical mechanisms to the spectral shape of variable blazars, as mentioned previously. It is 
indeed plausible that the spectrum could be described by a combination of different processes and not just a single 
electron distribution as assumed in each model. In particular, spectral curvature may be a combination of both a curved 
electron distribution and Klein-Nishina effects at the same time. It is also possible for the 
Klein-Nishina effects to affect the electron distribution \citep[e.g.,][]{moderskietal2005}. If the electron distribution 
is a power law which is hardened by inefficient Compton cooling at the high-energy end (if Compton cooling strongly 
dominates over other radiative cooling mechanisms), then this would result in a power law photon spectrum, which is 
inconsistent with most {\it Fermi}-LAT spectra investigated here.

The assumption of mono-energetic target photon spectra may also be an over-simplification, as broad non-thermal 
target photon distributions may result in additional spectral curvature \citep[see, e.g.,][]{tavecchioetal1998}. 
It is well known that Compton scattering of a broad non-thermal synchrotron spectrum by a broad non-thermal electron 
distribution introduces additional curvature, which is primarily caused by Klein-Nishina effects at high energies 
(which we are interested in here). As our results are well consistent with the standard paradigm that SSC dominates for 
BL Lacs, introducing the additional complication of SSC with a broad target photon spectrum would likely not yield any 
additional insights. For thermal target photon fields, however, the Compton spectrum is only weakly dependent on the 
distribution of seed photons, but depends critically on their characteristic energy, which is fitted within physically 
reasonable limits.

The best-fit values of $\nu_0 \sim 10^{16} - 10^{17}$ Hz for the KN model is compatible with the synchrotron emission 
from ISP and HSP blazars, thus favouring the SSC hypothesis. In the case of FSRQs, which are best fitted by a PL+EC in 
the Thomson regime, our results favour $\gamma$-ray emission scenarios based on Compton scattering of infrared 
radiation from the dust torus. This result is interesting for TeV telescopes since it will be possible for them to 
detect more FSRQs than if external photons were provided by the BLR. Indeed, very high energy measurements ($E>100$~GeV) 
with these telescopes, especially with the future Cherenkov Telescope Array, will help in characterizing the blazar 
$\gamma$-ray emission.

\acknowledgements{
The \textit{Fermi}-LAT Collaboration acknowledges generous ongoing support from a number of agencies and institutes 
that have supported both the development and the operation of the LAT as well as scientific data analysis. These 
include the National Aeronautics and Space Administration and the Department of Energy in the United States, the 
Commissariat \`a l'Energie Atomique and the Centre National de la Recherche Scientifique / Institut National de 
Physique Nucl\'eaire et de Physique des Particules in France, the Agenzia Spaziale Italiana and the Istituto Nazionale 
di Fisica Nucleare in Italy, the Ministry of Education, Culture, Sports, Science and Technology (MEXT), High Energy 
Accelerator Research Organization (KEK) and Japan Aerospace Exploration Agency (JAXA) in Japan, and the 
K.~A.~Wallenberg Foundation, the Swedish Research Council and the Swedish National Space Board in Sweden.
 
Additional support for science analysis during the operations phase is gratefully acknowledged from the Istituto 
Nazionale di Astrofisica in Italy and the Centre National d'\'Etudes Spatiales in France. This work performed in part 
under DOE Contract DE-AC02-76SF00515.

The authors thank Markos Georganopoulos and the anonymous reviewer for useful comments. JPvdB acknowledges the support of the National 
Astrophysics and Space Science Program (NASSP) in South Africa. MB acknowledges support through the South African Research Chair 
Initiative (SARChI) of the Department of Science and Technology and the National Research Foundation\footnote{Any 
opinion, finding and conclusion or recommendation expressed in this material is that of the authors, and the NRF does 
not accept any liability in this regard.} (NRF) of South Africa under NRF SARChI Chair grant no.~64789. AD thanks the 
support of the Juan de la Cierva program from the Spanish MEC.}


\begin{thebibliography}{}

\bibitem[Aartsen et al.(2018a)]{Aartsen18a}Aartsen, M. G., et al., 2018a, Science, 361, 1378

\bibitem[Aartsen et al.(2018b)]{Aartsen18b}Aartsen, M. G., et al., 2018b, Science, 361, 147

\bibitem[Abdo et al.(2010a)]{abdoetal2010a}Abdo, A.A., Ackermann, M., Ajello, M., et al. 2010a, ApJ, 710, 1271

\bibitem[Abdo et al.(2010b)]{abdoetal2010b}Abdo, A.A., Ackermann, M., Ajello, M., et al. 2010b, ApJ, 716, 30

\bibitem[Acero et al.(2015)]{aceroetal2015}Acero, F., Ackermann, M., Ajello, M., et al. 2015, ApJS, 218, 23A

\bibitem[Ackermann et al.(2016)]{ackermannetal2016}Ackermann, M., Ajello, M., Atwood, W.B., et al. 2016, ApJS, 222, 5

\bibitem[Ajello et al.(2017)]{ajello2017}Ajello, M., Atwood, W.B., Baldini, L., et al. 2017, ApJS, 232, 18

\bibitem[B\"ottcher(2007)]{boettcher2007}B\"ottcher, M., 2007, Astrophys. \& Space Science, 309, 95

\bibitem[B\"ottcher et al.(2012)]{bottcheretal2012}B\"{o}ttcher, M., Harris, D.E. \& Krawczynski, H. 2012, Extragalactic Jets from Active Galactic Nuclei (Chichester: Springer-Praxis)

\bibitem[B\"ottcher et al.(2013)]{bottcheretal2013}B\"ottcher, M., Reimer, A., Sweeney, K. \& Prakash, A. 2013, ApJ, 768, 54

\bibitem[Costamante et al.(2018)]{costamante2018}Costamante, L., Cutini, S., Tosti, G., et al. 2018, in prep.

\bibitem[Dermer \& Menon (2012)]{dermermenon2012}Dermer, C.D. \& Menon, G. 2009, High Energy Radiation from Black Holes: Gamma Rays, Cosmic Rays, and Neutrinos (New Jersey: Princeton University)

\bibitem[Dom\'{\i}nguez et al.(2011)]{dominguezetal2011}Dom\'{\i}nguez, A., Primack, J.R., Rosario, D.J., et al. 2011, MNRAS, 410, 2556

\bibitem[Dom\'{\i}nguez \& Ajello(2015)]{dominguez2015}Dom\'{\i}nguez, A. \& Ajello, M., 2015, ApJ, 813, L34

\bibitem[Ellison et al.(1990)]{ellisonetal1990}Ellison, D.C., Jones, F.C. \& Reynolds, S.P. 1990, ApJ, 360, 702

\bibitem[Fossati et al.(1998)]{fossatietal1998}Fossati, G., Maraschi, L., Celotti, A., Comastri, A., Ghisellini, G. 1998, MNRAS, 299, 433

\bibitem[Ghisellini et al.(2017)]{ghisellinietal2017}Ghisellini, G., Righi, C., Costamante, L. \& Tavecchio, F. 2017, MNRAS, 469, 255G

\bibitem[Gould \& Schr\'{e}der(1966)]{gouldschreder1966}Gould, R.J. \& Schr\'{e}der, G. 1966, PhRvL, 16, 252

\bibitem[Hauser \& Dwek(2001)]{hauserdwek2001}Hauser, M.G. \& Dwek, E. 2001, ARA\&A, 39, 249

\bibitem[Jones(1968)]{jones1968}Jones, F.C. 1968, Phys. Rev., 167, 1159

\bibitem[Katarzy\'n ski(2012)]{Katarzynski12}Katarzy\'n ski, K., 2012, A\&A, 537, A47

\bibitem[Kirk \& Heavens(1989)]{kirkheavens1989}Kirk, J.G. \& Heavens, A.F. 1989, MNRAS, 239, 995

\bibitem[Landau et al.(1986)]{landauetal1986}Landau, R., Golisch, B., Jones, T.J., et al. 1986, ApJ, 308, 78

\bibitem[Massaro et al.(2004)]{massaroetal2004}Massaro, E., Perri, M., Giommi, P., \& Nesci, R. 2004, A\&A, 413, 489

\bibitem[Massaro et al.(2006)]{massaroetal2006}Massaro, E., Tramacere, A., Perri, M., Giommi, P., \& Tosti, G. 2006, A\&A, 448, 861

\bibitem[Moderski et al.(2005)]{moderskietal2005}Moderski, R., Sikora, M., Coppi, P.S. \& Aharonian, F. 2005, MNRAS, 363, 954

\bibitem[Nikishov(1962)]{nikishov1962}Nikishov, A.I. 1962, Sov. Phys. JETP, 14, 393

\bibitem[Paliya et al.(2018)]{paliyaetal2018}Paliya, V.S., Zhang, H., B\"ottcher, M., Ajello, M., Dom\'inguez, A., Joshi, M., 
Hartmann, D., \& Stalin, C.S., 2018, ApJ, 863, 98

\bibitem[Summerlin \& Baring(2012)]{summerlinbaring2012}Summerlin, E.J. \& Baring, M.G. 2012, ApJ, 745, 63

\bibitem[Tavecchio et al.(1998)]{tavecchioetal1998}Tavecchio, F., Maraschi, L. \& Ghisellini, G. 1998, ApJ, 509, 608

\bibitem[Tramacere et al.(2007)]{tramacereetal2007}Tramacere, A., Massaro, F., \& Cavaliere, A. 2007, A\&A, 466, 521

\bibitem[Tramacere et al.(2011)]{tramacereetal2011}Tramacere, A., Massaro, E., \& Taylor, A.M. 2011, ApJ, 739, 66

\bibitem[Urry \& Padovani(1995)]{urrypadovani1995}Urry, C.M. \& Padovani, P. 1995, PASP, 107, 803
\end{thebibliography}
\end{document}